\documentclass[preprintnumbers,amsmath,amssymbm,prl]{revtex4}
\usepackage{epsfig}
\usepackage{graphicx}

\begin{document}
\title{Analytic treatment of the black-hole bomb}
\author{Shahar Hod}
\affiliation{The Ruppin Academic Center, Emeq Hefer 40250, Israel}
\affiliation{ }
\affiliation{The Hadassah Institute, Jerusalem 91010, Israel}
\author{Oded Hod}
\affiliation{The Sackler Faculty of Exact Sciences, Tel Aviv University, Tel Aviv 69978, Israel}
\date{\today}

\begin{abstract}
\ \ \ A bosonic field impinging on a rotating black hole can be
amplified as it scatters off the hole, a phenomena known as
superradiant scattering. If in addition the field has a non-zero rest
mass then the mass term effectively works as a mirror, reflecting the
scattered wave back towards the black hole. In this physical system,
known as a black-hole bomb, the wave may bounce back and forth between
the black hole and some turning point amplifying itself each
time. Consequently, the massive field grows exponentially over time
and is unstable. Former analytical estimations of the timescale
associated with the instability were restricted to the regimes
$M\mu\gg 1$ and $M\mu\ll 1$, where $M$ and $\mu$ are the masses of the
black hole and the field, respectively.  In these two limits the
growth rate of the field was found to be extremely weak. However,
subsequent numerical investigations have indicated that the
instability is actually greatest in the regime $M\mu=O(1)$, where the
previous analytical approximations break down. Thus, a new analytical
study of the instability timescale for the case $M\mu=O(1)$ is
physically well motivated. In this Letter we study {\it analytically}
for the first time the phenomena of superradiant instability (the
black-hole bomb mechanism) in this physically interesting regime --
the regime of greatest instability. We find an instability growth rate
of $\tau^{-1}\equiv\omega_I=1.7\times 10^{-3}M^{-1}$ for the fastest
growing mode. This instability is four orders of magnitude {\it
stronger} than has been previously estimated.
\end{abstract}
\bigskip
\maketitle

%]

Black holes are the fundamental ``atoms" of general relativity. They
also play a central role in high energy physics, astrophysics, and
even in condensed matter physics. The fundamental role of black
holes makes it highly important to study the nature of their
stability: If a black hole is perturbed in some small way, will the
perturbation die away over time? Or will it grow exponentially until
it can no longer be considered a perturbation and hence demonstrate
the instability of the black hole?

The issue of black hole stability was first addressed by Regge and
Wheeler \cite{RegWheel} who demonstrated the stability of the
spherically symmetric Schwarzschild black hole. If a Schwarzschild
black hole is perturbed, then the perturbation will oscillate and damp
out over time \cite{Vish}. This implies that perturbation fields would
either be radiated away to infinity or swallowed by the black hole.

The stability question of rotating Kerr black holes is a bit more
involved. Press and Teukolsky \cite{PressTeu1,Teu} have shown that
rotating black holes are stable under free gravitational
perturbations (see also \cite{Hodp1} and references therein).
However, the superradiance effect may change this conclusion.
Superradiant scattering is a well-known phenomena in quantum systems
\cite{Mano,Grein} as well as in classical ones \cite{Zel,Car}.
Considering a wave of the form $e^{im\phi}e^{-i\omega t}$ incident
upon a rotating object whose angular velocity is $\Omega$, one finds
that if the frequency $\omega$ of the incident wave satisfies the
relation
\begin{equation}\label{Eq1}
\omega<m\Omega\  ,
\end{equation}
then the scattered wave is amplified.

A bosonic field impinging upon a rotating Kerr black hole can be
amplified if the superradiance condition (\ref{Eq1}) is satisfied,
where in this case $\Omega={a\over {r^2_++a^2}}$ is the angular
velocity of the black-hole horizon. Here $r_+$ and $a$ are the
horizon radius and the angular momentum per unit mass of the black
hole, respectively. The energy radiated away to infinity may
actually exceed the energy present in the initial perturbation.
Feeding back the amplified scattered wave, one can gradually extract
the rotational energy of the black hole. Press and Teukolsky
suggested to use this mechanism to build a {\it black-hole bomb}
\cite{PressTeu2}: If one surrounds the black hole by a reflecting
mirror, the wave will bounce back and forth between the black hole
and the mirror amplifying itself each time. Thus, the total energy
extracted from the black hole will gradually grow.

Remarkably, nature sometimes provides its own mirror \cite{Car}: If
one considers a {\it massive} scalar field with mass ${\cal M}$
scattered off a rotating black hole, then for $\omega<\mu\equiv
{\cal M}G/\hbar c$ the mass term effectively works as a mirror
\cite{Dam,Zour,Det,Furu,Dolan}. The physical idea is to consider a
wave packet of the massive field in a bound orbit around the black
hole \cite{Zour,Det}. The gravitational force binds the field and
keeps it from escaping to infinity. At the event horizon some of the
field goes down the black hole, and if the frequency of the wave is
in the superradiance regime (\ref{Eq1}) then the field is amplified.
In this way the field is amplified at the horizon while being bound
away from infinity. Consequently, the massive field grows
exponentially over time and is unstable \cite{Det}.

The nature of the superradiant instability (the black hole bomb)
depends on two parameters:
\begin{itemize}
\item{The rotation rate $a$ of the black hole.}
\item{The dimensionless product of the black hole mass $M$ and the field
mass $\mu$. The product $M\mu$ is actually the ratio of the black
hole size to the Compton wavelength associated with the rest mass of
the field. (we shall henceforth use natural units in which
$G=c=\hbar=1$. In these units $\mu$ has the dimensions of
$1/$length.)}
\end{itemize}

Former analytical estimates of the instability timescale associated
with the dynamics a of massive scalar field in the rotating Kerr
spacetime were restricted to the regimes $M\mu\gg 1$ \cite{Zour} and
$M\mu\ll 1$ \cite{Det,Furu}. In these two limits the growth rate of
the field (the imaginary part $\omega_I$ of the mode's frequency) was
found to be very weak, scaling like $M^{-1}e^{-1.84M\mu}\ll 1$ for
$M\mu\gg 1$ \cite{Zour} and like $M^{-1}(M\mu)^9\ll 1$ for the
$M\mu\ll 1$ case \cite{Det,Furu}. We note, however, that the former
analytical approximations \cite{Zour,Det,Furu} fail in the regime
$M\mu=O(1)$. Thus, direct {\it numerical} integration of the
perturbation equations seemed necessary to find the actual growth rate
of the perturbations in this regime \cite{Zour,Furu,Dolan}. These
numerical investigations have indicated that the superradiant
instability is in fact greatest in the regime $M\mu=O(1)$
\cite{Zour,Furu,Dolan}. A new {\it analytical} study of the
superradiant instability in the regime $M\mu=O(1)$ is therefore
physically desirable.

It is worth noting that, previous numerical investigations
\cite{Zour,Furu,Dolan} of the black-hole bomb have indicated that
the instability is most effective under the following conditions:
\begin{itemize}
\item{The black hole is maximally rotating with $a\simeq M$.}
\item{The dimensionless product $M\mu$ satisfies the relation $M\mu\sim {1\over 2}$.}
\item{The frequency of the unstable mode satisfies the relations
$\omega\simeq m\Omega$ and $\omega\simeq\mu$. (Of course, for the
mode to be in the superradiant regime one should have
$\omega<m\Omega$. In addition, for the mode to be in a bound state
it should satisfy $\omega<\mu$.)}
\end{itemize}

As we shall show below, the black-hole bomb and the associated
instability timescale can be studied {\it analytically} in the above
mentioned regime of physical interest (the regime of the greatest
instability). The physical system we consider consists of a massive
scalar field coupled to a rotating Kerr black hole. The dynamics of
a scalar field $\Psi$ of mass $\mu$ in the Kerr spacetime
\cite{Kerr} is governed by the Klein-Gordon equation
\begin{equation}\label{Eq2}
(\nabla^a \nabla_a -\mu^2)\Psi=0\  .
\end{equation}
One may decompose the field as
\begin{equation}\label{Eq3}
\Psi_{lm}(t,r,\theta,\phi)=e^{im\phi}S_{lm}(\theta;a\omega)R_{lm}(r;a\omega)e^{-i\omega
t}\ ,
\end{equation}
where $(t,r,\theta,\phi)$ are the Boyer-Lindquist coordinates
\cite{Kerr}, $\omega$ is the (conserved) frequency of the mode, $l$
is the spheroidal harmonic index, and $m$ is the azimuthal harmonic
index with $-l\leq m\leq l$. (We shall henceforth omit the indices
$l$ and $m$ for brevity.) With the decomposition (\ref{Eq3}), $R$
and $S$ obey radial and angular equations both of confluent Heun
type coupled by a separation constant $A(a\omega)$ \cite{Heun,Flam}.
The sign of $\omega_I$ determines whether the solution is decaying
$(\omega_I<0)$ or growing $(\omega_I>0)$ in time.

The angular functions $S(\theta;a\omega)$ are the spheroidal
harmonics which are solutions of the angular equation
\cite{Teu,Flam}
\begin{equation}\label{Eq4}
{1\over {\sin\theta}}{\partial \over
{\partial\theta}}\Big(\sin\theta {{\partial
S}\over{\partial\theta}}\Big)+\Big[a^2(\omega^2-\mu^2)\cos^2\theta-{{m^2}\over{\sin^2\theta}}+A\Big]S=0\
.
\end{equation}
The angular functions are required to be regular at the poles
$\theta=0$ and $\theta=\pi$. These boundary conditions pick out a
discrete set of eigenvalues $\{A_l\}$ labeled by an integer $l$. For
$\omega\simeq\mu$ one can treat $a^2(\omega^2-\mu^2)\cos^2\theta$ in
Eq. (\ref{Eq4}) as a perturbation term on the generalized Legendre
equation. We can then expand the separation constants in powers of
$a^2(\mu^2-\omega^2)$ to find \cite{Abram}
\begin{equation}\label{Eq5}
A=l(l+1)+\sum_{k=1}^{\infty}c_ka^{2k}(\mu^2-\omega^2)^k\  .
\end{equation}
The expansion coefficients $\{c_k\}$ are given in \cite{Abram}. [For
example, for $l=m=1$, the case of physical interest (see below), we
find $c_1=-4/5,\ c_2=-4/875,\ c_3=8/65625,...$].

The radial Teukolsky equation is given by \cite{Teuk,Hodcen,Stro}
\begin{equation}\label{Eq6}
\Delta{{d} \over{dr}}\Big(\Delta{{dR}\over{dr}}\Big)+\Big[K^2
-\Delta(a^2\omega^2-2ma\omega+\mu^2r^2+A)\Big]R=0\ ,
\end{equation}
where $\Delta\equiv r^2-2Mr+a^2$ and $K\equiv (r^2+a^2)\omega-am$.
The zeroes of $\Delta$, $r_{\pm}=M\pm (M^2-a^2)^{1/2}$, are the
black hole (event and inner) horizons.

We are interested in solutions of the radial equation (\ref{Eq6})
with the physical boundary conditions of purely ingoing waves at the
black-hole horizon (as measured by a comoving observer) and a
decaying (bounded) solution at spatial infinity \cite{Zour,Dolan}.
That is,

\begin{equation}\label{Eq7}
%\label{eq:boundary_conditions}
R \sim
\begin{cases}
{1\over r}e^{-\sqrt{\mu^2-\omega^2}y} & \text{ as }
r\rightarrow\infty\ \ (y\rightarrow \infty)\ ; \\
e^{-i (\omega-m\Omega)y} & \text{ as } r\rightarrow r_+\ \
(y\rightarrow -\infty)\ ,
\end{cases}
\end{equation}
%\begin{equation}\label{Eq4}
%\psi(r\to r_+)\sim {\cal T}(\omega)e^{-i(\omega-m\Omega)y}\ \ \ ;\ \
%\ \psi(r\to\infty) \sim e^{-i\omega r}+{\cal R}(\omega)e^{i\omega
%r}\ .
%\end{equation}
where the ``tortoise" radial coordinate $y$ is defined by
$dy=[(r^2+a^2)/\Delta]dr$.  These boundary conditions single out a
discrete set of resonances $\{\omega_n\}$ which correspond to bound
states of the massive field \cite{Zour,Dolan}.

It is convenient to define new dimensionless variables
\begin{equation}\label{Eq8}
x\equiv {{r-r_+}\over {r_+}}\ \ ;\ \ \tau\equiv{{r_+-r_-}\over
{r_+}}\ \ ;\ \ \varpi\equiv{{\omega-m\Omega}\over{2\pi T_{BH}}}\ \
;\ \ k\equiv 2\omega r_+\  ,
\end{equation}
in terms of which the radial equation becomes
\begin{equation}\label{Eq9}
x(x+\tau){{d^2R}\over{dx^2}}+(2x+\tau){{dR}\over{dx}}+VR=0\  ,
\end{equation}
where $V\equiv
K^2/r^2_+x(x+\tau)-[a^2\omega^2-2ma\omega+\mu^2r^2_+(x+1)^2+A]$ and
$K=r^2_+\omega x^2+r_+kx+r_+\varpi\tau/2$.

As discussed above, previous numerical investigations
\cite{Zour,Furu,Dolan} have indicated that the black-hole
instability is most pronounced in the regime $\tau\ll 1$ with
$M(m\Omega-\omega)\ll 1$. As we shall now show, the radial equation
is amenable to an analytic treatment in this regime of physical
interest.

We first consider the radial equation (\ref{Eq9}) in the far region
$x\gg \text{max}\{\tau,M(m\Omega-\omega)\}$. Then Eq. (\ref{Eq9}) is
well approximated by
\begin{equation}\label{Eq10}
x^2{{d^2R}\over{dx^2}}+2x{{dR}\over{dx}}+V_{\text{far}}R=0\  ,
\end{equation}
where $V_{\text{{far}}}=(\omega^2-\mu^2)r^2_+x^2+2(\omega
k-\mu^2r_+)r_+x-(a^2\omega^2-2ma\omega+\mu^2r^2_++A-k^2)$. A
solution of Eq. (\ref{Eq10}) that satisfies the boundary condition
(\ref{Eq7}) can be expressed in terms of the confluent
hypergeometric functions $M(a,b,z)$ \cite{Morse,Abram}
\begin{equation}\label{Eq11}
R=C_1(2\sqrt{\mu^2-\omega^2}r_+)^{{1\over 2}+\beta}x^{-{1\over
2}+\beta}e^{-\sqrt{\mu^2-\omega^2}r_+x}M({1\over
2}+\beta-\kappa,1+2\beta,2\sqrt{\mu^2-\omega^2}r_+x)+C_2(\beta\to
-\beta)\  ,
\end{equation}
where $C_1$ and $C_2$ are constants and
\begin{equation}\label{Eq12}
\beta^2\equiv a^2\omega^2-2ma\omega+\mu^2r^2_++A-k^2+{1\over 4}\ \
;\ \ \kappa\equiv {{\omega k-\mu^2r_+}\over{\sqrt{\mu^2-\omega^2}}}\
.
\end{equation}
The notation $(\beta\to -\beta)$ means ``replace $\beta$ by $-\beta$
in the preceding term."

We next consider the near horizon region $x\ll 1$. The radial
equation is given by Eq. (\ref{Eq9}) with $V\to
V_{\text{near}}\equiv-(a^2\omega^2-2ma\omega+\mu^2r^2_++A)+(kx+\varpi\tau/2)^2/x(x+\tau)$.
The physical solution obeying the ingoing boundary conditions at the
horizon is given by \cite{Morse,Abram}
\begin{equation}\label{Eq13}
R=x^{-{i\over 2}\varpi}\Big({x\over \tau}+1\Big)^{i({1\over
2}\varpi-k)}{_2F_1}({1\over 2}+\beta-ik,{1\over
2}-\beta-ik;1-i\varpi;-x/\tau)\  ,
\end{equation}
where $_2F_1(a,b;c;z)$ is the hypergeometric function.

The solutions (\ref{Eq11}) and (\ref{Eq13}) can be matched in the
overlap region $\text{max}\{\tau,M(m\Omega-\omega)\}\ll x\ll 1$. The
$x\ll 1$ limit of Eq. (\ref{Eq11}) yields \cite{Morse,Abram}
\begin{equation}\label{Eq14}
R\to C_1(2\sqrt{\mu^2-\omega^2}r_+)^{{1\over 2}+\beta}x^{-{1\over
2}+\beta}+C_2(\beta\to -\beta)\  .
\end{equation}
The $x\gg \tau$ limit of Eq. (\ref{Eq13}) yields \cite{Morse,Abram}
\begin{equation}\label{Eq15}
R\to \tau^{{1\over
2}-\beta-i\varpi/2}{{\Gamma(2\beta)\Gamma(1-i\varpi)}\over{\Gamma({1\over
2}+\beta-ik)\Gamma({1\over 2}+\beta-i\varpi+ik)}}x^{-{1\over
2}+\beta}+(\beta\to -\beta)\  .
\end{equation}
By matching the two solutions in the overlap region one finds
\begin{equation}\label{Eq16}
C_1=\tau^{{1\over
2}-\beta-i\varpi/2}{{\Gamma(2\beta)\Gamma(1-i\varpi)}\over{\Gamma({1\over
2}+\beta-ik)\Gamma({1\over
2}+\beta-i\varpi+ik)}}(2\sqrt{\mu^2-\omega^2}r_+)^{-{1\over
2}-\beta}\  ,
\end{equation}
\begin{equation}\label{Eq17}
C_2=\tau^{{1\over
2}+\beta-i\varpi/2}{{\Gamma(-2\beta)\Gamma(1-i\varpi)}\over{\Gamma({1\over
2}-\beta-ik)\Gamma({1\over
2}-\beta-i\varpi+ik)}}(2\sqrt{\mu^2-\omega^2}r_+)^{-{1\over
2}+\beta}\  .
\end{equation}

Approximating Eq. (\ref{Eq11}) for $x\to\infty$ one gets
\cite{Morse,Abram}
\begin{eqnarray}\label{Eq18}
R&\to&
\Big[C_1(2\sqrt{\mu^2-\omega^2}r_+)^{-\kappa}{{\Gamma(1+2\beta)}\over{\Gamma({1\over
2}+\beta-\kappa)}}x^{-1-\kappa}+C_2(\beta\to
-\beta)\Big]e^{\sqrt{\mu^2-\omega^2}r_+x}\nonumber
\\&& + \Big[C_1(2\sqrt{\mu^2-\omega^2}r_+)^{\kappa}{{\Gamma(1+2\beta)}\over{\Gamma({1\over
2}+\beta+\kappa)}}x^{-1+\kappa}(-1)^{-{1\over
2}-\beta+\kappa}+C_2(\beta\to
-\beta)\Big]e^{-\sqrt{\mu^2-\omega^2}r_+x}\ .
\end{eqnarray}
A bound state is characterized by a decaying field at spatial
infinity. The coefficient of the growing exponent
$e^{\sqrt{\mu^2-\omega^2}r_+x}$ in Eq. (\ref{Eq18}) should therefore
vanish. Taking cognizance of Eqs. (\ref{Eq16})-(\ref{Eq18}) for
$\varpi\gg 1$, one finds the resonance condition for the bound
states of the field
\begin{equation}\label{Eq19}
{1\over{\Gamma({1\over
2}+\beta-\kappa)}}=(8i)^{2\beta}\Big[{{\Gamma(-2\beta)}\over{\Gamma(2\beta)}}\Big]^2{{\Gamma({1\over
2}+\beta-ik)}\over{\Gamma({1\over 2}-\beta-ik)\Gamma({1\over
2}-\beta-\kappa)}}\Big[M^2\sqrt{\mu^2-\omega^2}(m\Omega-\omega)\Big]^{2\beta}\
.
\end{equation}

The growing resonances of the field can be estimated analytically in
the regime of physical interest $\omega\simeq\mu\simeq m\Omega$: We
first note that in this regime the right-hand side (RHS) of Eq.
(\ref{Eq19}) is small [due to the factors $M\sqrt{\mu^2-\omega^2}$
and $M(m\Omega-\omega)$]. One may therefore write a zeroth-order
approximation for the resonance condition: $\text{LHS}\equiv
1/\Gamma({1\over 2}+\beta-\kappa)\simeq 0$. Using the well-known
pole structure of the Gamma functions \cite{Abram}, one finds the
approximated resonance condition
\begin{equation}\label{Eq20}
{1\over 2}+\beta-\kappa=-n  ,
\end{equation}
where $n\geq 0$ is a non-negative integer. Taking cognizance of Eqs.
(\ref{Eq5}) and (\ref{Eq12}), one realizes that Eq. (\ref{Eq20}) is
a simple polynomial equation for the variable $M^2(\mu^2-\omega^2)$,
whose solutions we denote by $\omega^{(0)}_R$. [Note that for the
zeroth-order approximation one has $\omega^{(0)}_I=0$.] One can then
use $\omega^{(0)}_R$ in the equation
$\Im\text{LHS}(\omega^{(0)}_R,\omega^{(1)}_I)=
\Im\text{RHS}(\omega^{(0)}_R,\omega^{(1)}_I)$ to obtain a simple
polynomial equation for the first-order solution $\omega^{(1)}_I$.
%The
%analytical approximation can be further improved using an iteration
%scheme: One may solve the equation
%$\Re\text{LHS}(\omega^{(1)}_R,\omega^{(1)}_I)=\Re\text{RHS}(\omega^{(0)}_R,\omega^{(1)}_I)$
%[this is again a polynomial equation for the variable
%$M^2(\mu^2-\omega^2)$] to find the first-order solution
%$\{\omega^{(1)}_R\}$. This solution can then be used to evaluate
%$\{\omega^{(2)}_I\}$ using the polynomial equation
%$\Im\text{LHS}(\omega^{(1)}_R,\omega^{(2)}_I)=
%\Im\text{RHS}(\omega^{(1)}_R,\omega^{(2)}_I)$. This simple iteration
%scheme can be repeated as many times as desired.

\begin{figure}[h]
  \begin{center}
    \epsfxsize=10.0cm \epsffile{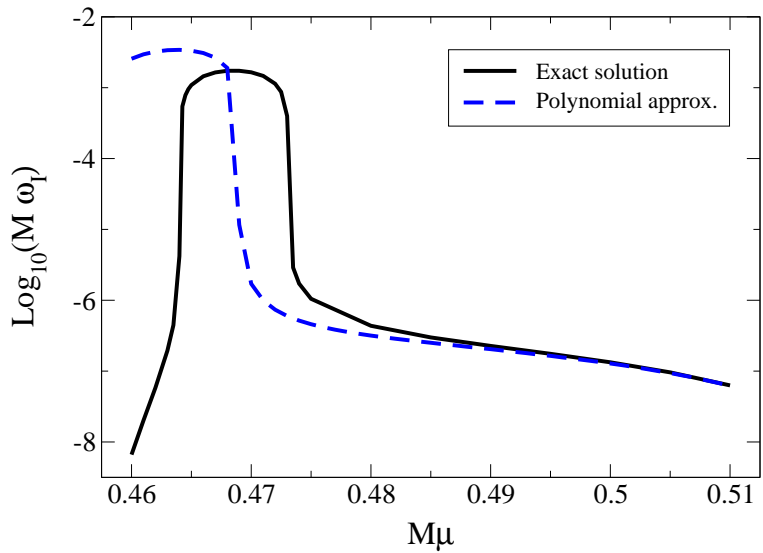}
  \end{center}
  \caption{Superradiant instability for maximally rotating black holes
    $(a\simeq M)$. The results are for the $l=m=1$ mode, the mode with the
    greatest instability. The growth rate $M\omega_I$ is shown as a
    function of the dimensionless product $M\mu$. We display results for
    the direct solutions of both the exact resonance condition
    (\ref{Eq19}) and the polynomial approximation (\ref{Eq20}). The
    maximum growth rate is $\tau^{-1}\equiv\omega_I=1.7\times
    10^{-3}M^{-1}$, where $\tau$ is the $e$-folding time.}
  \label{Fig1}
\end{figure}

In figure \ref{Fig1} we depict results for the most unstable
(fastest growing) mode with $l=m=1$. We present results of the
direct solutions of both the exact resonance condition (\ref{Eq19})
and the polynomial approximation (\ref{Eq20}). One finds a good
qualitative agreement between the two. The maximum growth rate we
find is $\tau^{-1}\equiv\omega_I=1.7\times 10^{-3}M^{-1}$, where
$\tau$ is the $e$-folding time. We would like to emphasize that,
this growth rate is four orders of magnitude stronger than has been
previously found.

\input{epsf}

What are the observable consequences of this instability? Let us
consider for example the neutral spinless pion $\pi^0$ whose mass is
$\mu\simeq 134.97$\ MeV/c$^2$. The superradiant instability is most
pronounced for $M\mu\simeq 0.469$, which corresponds to a primordial
black hole of mass $M\simeq 9.3\times 10^{11}$\ Kg \cite{Dolan}. For
the superradiant instability to be effective, the lifetime of the
neutral pion, $\tau_{1/2}\simeq 8.2\times 10^{-17}$sec, should be
significantly longer than the timescale associated with the
instability, $\tau=(1.7\times 10^{-3})^{-1}GM/c^3\simeq 1.3\times
10^{-21}$sec. This condition is indeed satisfied by more than four
orders of magnitude. One therefore concludes that the superradiant
instability in the neutral pion channel may indeed manifest itself
for primordial black holes.

In summary, we have studied analytically the instability of rotating
black holes to perturbations of massive scalar fields. Former
analytical estimations \cite{Zour,Det,Furu} of the timescale
associated with the instability were restricted to the regimes
$M\mu\gg 1$ and $M\mu\ll 1$. In these two limits the growth rate of
the field was found to be extremely weak. However, subsequent
numerical investigations \cite{Zour,Furu,Dolan} have indicated that
the instability is actually greatest in the regime $M\mu=O(1)$, where
unfortunately the previous analytical approximations are not suitable
to describe the dynamics of this instability. Motivated by these
numerical studies, we have provided here for the first time an
analytic treatment of the superradiant instability (the black-hole
bomb mechanism) in the physically most interesting regime $M\mu\sim
{1\over 2}$, where the instability is most pronounced. We find an
instability growth rate of $\tau^{-1}\equiv\omega_I=1.7\times
10^{-3}M^{-1}$ for the fastest growing mode-- four orders of magnitude
{\it stronger} than has been previously estimated.

\bigskip
\noindent
{\bf ACKNOWLEDGMENTS}
\bigskip

This research is supported by the Meltzer Science Foundation. We thank
Liran Shimshi, Clovis Hopman, Yael Oren, Adi Zalckvar, Ophir Ariel,
and Arbel M. Ongo for helpful discussions.

%\newpage

\end{document}